\documentclass[twocolumn,prl,showpacs,amsmath,amssymb]{revtex4}

\usepackage{graphicx}% Include figure files
\usepackage{dcolumn}% Align table columns on decimal point
\usepackage{bm}% bold math

\begin{document}

%\preprint{TU/01}
\title{Measurement of the Generalized Forward Spin Polarizabilities of the Neutron}

\author{
M.~Amarian$^{\nyerevan}$,
L.~Auerbach$^{\ntemple}$,
T.~Averett$^{\njlab,\nwm}$,
J.~Berthot$^{\nclermont}$,
P.~Bertin$^{\nclermont}$,
W.~Bertozzi$^{\nmit}$,
T.~Black$^{\nmit}$,
E.~Brash$^{\nregina}$,
D.~Brown$^{\nmaryland}$,
E.~Burtin$^{\nsaclay}$
% ...
}
\affiliation{
\baselineskip 2 pt
\vskip 0.3 cm
{\rm (Jefferson Lab E94010 Collaboration)} \break
\vskip 0.1 cm
\centerline{{$^{\ncaltech}$California Institute of Technology, Pasadena, California 91125}}
\centerline{{$^{\nchungham}$Chungnam National University, Taejon 305-764, Korea}}
\centerline{{$^{\nhampton}$Hampton University, Hampton, Virginia 23668}}
\centerline{{$^{\nclermont}$LPC IN2P3/CNRS, Universit\'e Blaise Pascal, F--63170 Aubi\`ere Cedex,
France}}
\centerline{{$^{\ninfn}$Istituto Nazionale di Fiscica Nucleare, Sezione Sanit\`a, 00161 Roma, Italy}}
\centerline{{$^{\njlab}$Thomas Jefferson National Accelerator Facility, Newport News, Virginia
23606}} 
\centerline{{$^{\nkent}$Kent State University, Kent, Ohio 44242}}
\centerline{{$^{\nkentucky}$University of Kentucky, Lexington, Kentucky 40506}}
\centerline{{$^{\nkharkov}$Kharkov Institute of Physics and Technology, Kharkov 310108, Ukraine}}
\centerline{{$^{\nmaryland}$University of Maryland, College Park, Maryland 20742}}
\centerline{{$^{\nmit}$Massachusetts Institute of Technology, Cambridge, Massachusetts 02139}}
\centerline{{$^{\numass}$University of Massachusetts-Amherst, Amherst, Massachusetts 01003}}
\centerline{{$^{\nunh}$University of New Hampshire, Durham, New Hampshire 03824}}
\centerline{{$^{\nodu}$Old Dominion University,  Norfolk, Virginia 23529}}
\centerline{{$^{\nprinceton}$Princeton University, Princeton, New Jersey 08544}}
\centerline{{$^{\nregina}$University of Regina, Regina, SK S4S 0A2, Canada}}
\centerline{{$^{\nrutgers}$Rutgers, The State University of New Jersey, Piscataway, New Jersey
08855}}
\centerline{{$^{\nsaclay}$CEA Saclay, DAPNIA/SPHN, F--91191 Gif/Yvette, France}}
\centerline{{$^{\nsyracuse}$Syracuse University, Syracuse, New York 13244}}
\centerline{{$^{\ntemple}$Temple University, Philadelphia, Pennsylvania 19122}}
\centerline{{$^{\ntohoku}$Tohoku University, Sendai 980, Japan}}
\centerline{{$^{\nuva}$University of Virginia, Charlottesville, Virginia 22904}}
\centerline{{$^{\nwm}$The College of William and Mary, Williamsburg, Virginia 23187}}
\centerline{{$^{\nyerevan}$Yerevan Physics Institute, Yerevan 375036, Armenia}}
}

\newcommand{\ncaltech}{1}
\newcommand{\nchungham}{2}
\newcommand{\nhampton}{3}
\newcommand{\nclermont}{4}
\newcommand{\ninfn}{5}
\newcommand{\njlab}{6}
\newcommand{\nkent}{7}
\newcommand{\nkentucky}{8}
\newcommand{\nkharkov}{9}
\newcommand{\nmaryland}{10}
\newcommand{\nmit}{11}
\newcommand{\numass}{12}
\newcommand{\nunh}{13}
\newcommand{\nodu}{14}
\newcommand{\nprinceton}{15}
\newcommand{\nregina}{16}
\newcommand{\nrutgers}{17}
\newcommand{\nsaclay}{18}
\newcommand{\nsyracuse}{19}
\newcommand{\ntemple}{20}
\newcommand{\ntohoku}{21}
\newcommand{\nuva}{22}
\newcommand{\nwm}{23}
\newcommand{\nyerevan}{24}

\date{\today}

\begin{abstract}
The generalized forward
spin polarizabilities $\gamma_0$ and $\delta_{LT}$ of the neutron
have been extracted for the first time 
in a $Q^2$ range from 0.1 to 0.9 GeV$^2$.  
Since $\gamma_0$ is sensitive to nucleon resonances and 
$\delta_{LT}$ is insensitive to the $\Delta$ resonance, it is expected 
that the pair of forward spin polarizabilities 
should provide benchmark tests of the current understanding of the chiral 
dynamics of QCD.
The new results on $\delta_{LT}$ show   
significant disagreement with
Chiral Perturbation Theory calculations, while the data for $\gamma_0$
at low $Q^2$ are in good agreement with a next-to-lead order Relativistic 
Baryon Chiral Perturbation theory calculation.
The data show good agreement with the phenomenological MAID model.

\end{abstract}


\maketitle

The study of nucleon structure is one of the most important subjects of modern 
physics. The dominant interaction responsible for nucleon structure is the
strong interaction. High energy experiments have established Quantum 
Chromodynamics (QCD) as the gauge theory describing the strong interaction
between quarks and gluons, which are the elementary constituents of the
nucleon. At high energies, observables in QCD can be calculated 
perturbatively since the running coupling constant is small. 
However, at low energies, the coupling constant becomes increasingly large and
quarks and gluons are confined to color singlet objects known as hadrons. 
There, 
nucleon structure is usually described in terms of hadronic degrees of 
freedom, namely baryons and mesons. Chiral Perturbation Theory ($\chi$PT) is 
applicable in this region.
The intermediate energy region is described by phenomenological models
and will eventually be 
described by Lattice QCD calculations.

The polarizabilities of the nucleon are fundamental observables 
that characterize  
nucleon structure. They are related to integrals of the nucleon
excitation spectrum. 
The electric and magnetic polarizabilities measure the nucleon's 
response to an external electromagnetic field.
Because the polarizabilities can be linked to the forward Compton scattering
amplitudes,
real photon Compton scattering experiments~\cite{RCS} 
were performed to measure these polarizabilities.
Another polarizability, associated with a spin-flip, is the forward spin 
polarizability $\gamma_0$. It has been measured in an experiment at MAMI 
(Mainz)~\cite{mainz} with a circularly polarized photon beam on a 
longitudinally polarized proton target.
The extension of these quantities to the case of virtual photon Compton scattering 
with finite four-momentum-squared, $Q^2$, leads to the concept of the
generalized polarizabilities~\cite{GPT}.
Generalized polarizabilities are related to the forward
virtual Compton scattering (VCS) amplitudes and the forward doubly-virtual
Compton scattering (VVCS) amplitudes~\cite{DPV}. 
With this additional dependence on $Q^2$, 
the generalized polarizabilities provide a powerful tool to probe the nucleon 
structure covering the whole range from the partonic to the hadronic region. 
In particular,
the generalized polarizabilities provide one of the most extensive tests 
of 
$\chi$PT calculations in the low $Q^2$ region~\cite{DPV,GPE}. However, up to 
now, other than the real photon measurement of $\gamma_0$ for the proton from 
MAMI, there are no experimental data available for the generalized spin 
polarizabilities for either the proton or the neutron.

In this paper, we present the first results for the neutron generalized forward
spin polarizabilities $\gamma_0(Q^2)$ and $\delta_{LT}(Q^2)$
over the $Q^2$ range from 0.1 to 0.9 (GeV)$^2$.
These results were extracted from a measurement of $\sigma_{TT}$ and 
$\sigma_{LT}$, 
the doubly polarized transverse-transverse and  
longitudinal-transverse interference cross sections,
or equivalently $g_1$ and $g_2$, the two inclusive spin structure 
functions, in the resonance region.
Jefferson Lab's high intensity polarized electron beam and a high 
density polarized $^3$He target were used for the measurement. The polarized 
$^3$He target provided an effective polarized neutron target
because the ground state of $^3$He is dominated by the $s$ state, in which the
spins of the two protons anti-align and cancel. Therefore the spin of the 
$^3$He nucleus comes largely from the neutron.
Doubly polarized inclusive cross sections were 
measured at six incident beam energies from 0.86 to 5.1 GeV, all at a fixed
scattering angle of 15.5$^\circ$. Data were collected
for both longitudinal and transverse target polarization orientations, enabling
the extractions of both $\sigma_{TT}$ and $\sigma_{LT}$.
The integrals of
$\sigma_{TT}$ and $\sigma_{LT}$ of the neutron were extracted from those of the
$^3$He following the prescription suggested by Ciofi degli Atti and Scopetta in
Ref.~\cite{ciofi} to take into account the nuclear corrections. 
Details of the experiment can be found in Refs.~\cite{Ama:02,Ama:03}.

Following Ref. \cite{DPV}, an unsubtracted dispersion relation for
the spin-flip VVCS amplitude $g_{TT}$ with an appropriate convergence
behavior at high-energy leads to

\begin{equation}
{\rm Re}~\tilde{g}_{TT}(\nu,Q^2)
=
(\frac{\nu}{2 \pi^2}){\cal P}\int^{\infty}_{\nu_0}\frac{K(\nu',Q^2)
\sigma_{TT}(\nu',Q^2)}{\nu'^2-\nu^2}d\nu', 
\end{equation}
where $\tilde{g}_{TT} \equiv g_{TT} - g_{TT}^{pole}$, $g_{TT}^{pole}$ is the 
nucleon pole contribution,
$\nu$ is the energy of the virtual photon and
$K$ is the virtual photon flux factor.
The lower limit of the integration $\nu_0$ is the $\pi$ production threshold on
the neutron.
A low energy expansion gives:
\begin{equation}
{\rm Re}~\tilde{g}_{TT}(\nu,Q^2)=
(\frac{2\alpha}{M^2})I_A(Q^2)\nu+\gamma_0(Q^2)\nu^3+O(\nu^5),
\end{equation}
with $\alpha$ the electromagnetic fine-structure constant and $M$ the 
neutron mass. $I_A(Q^2)$ is the coefficient of the $O(\nu)$ term of
the Compton amplitude.
Equation (2) defines the generalized forward spin 
polarizability $\gamma_0(Q^2)$.
Combining Eqs. (1) and (2), the $O(\nu)$ term yields a sum rule 
for the generalized Gerasimov-Drell-Hearn (GDH) integral~\cite{GDH, GGDH1,GGDH2}:
the integration of $\sigma_{TT}$, with $1/\nu$ weighting, is proportional to
$I_A$, the coefficient of the $O(\nu)$ term of the VVCS amplitude.
From the $O(\nu^3)$ term, one obtains a sum 
rule for the generalized forward spin polarizability~\cite{DPV}:
\begin{eqnarray}
\gamma_0(Q^2)&=&
(\frac{1}{2\pi^2})\int^{\infty}_{\nu_0}\frac{K(\nu,Q^2)}{\nu}
\frac{\sigma_{TT}(\nu,Q^2)}{\nu^3}d\nu \\
=&&\hspace{-0.5cm}\frac{16 \alpha M^2}{Q^6}\int^{x_0}_0 x^2 \Bigl [g_1(x,Q^2)-\frac{4M^2}{Q^2}
x^2g_2(x,Q^2)\Bigr ] dx, \nonumber
\end{eqnarray}
where $x=Q^2/(2M\nu)$ is the Bjorken scaling variable. 

Considering the longitudinal-transverse interference amplitude $g_{LT}$,
with the same assumptions, one obtains:
\begin{equation}
{\rm Re}~\tilde{g}_{LT}(\nu,Q^2)=
(\frac{2\alpha}{M^2})Q I_3(Q^2)+Q \delta_{LT}(Q^2)\nu^2+O(\nu^4) 
\end{equation}
where the $O(1)$ term leads to a sum rule for $I_3(Q^2)$, which relates it to
the $\sigma_{LT}$ integral over the excitation spectrum.
The $O(\nu^2)$ term leads to the generalized  longitudinal-transverse
polarizability~\cite{DPV}:
\begin{eqnarray}
\delta_{LT}(Q^2)&=&
(\frac{1}{2\pi^2})\int^{\infty}_{\nu_0}\frac{K(\nu,Q^2)}{\nu}
\frac{\sigma_{LT}(\nu,Q^2)}{Q \nu^2}d\nu \\
=&&\hspace{-0.5cm}\frac{16 \alpha M^2}{Q^6}\int^{x_0}_0 x^2 \Bigl [g_1(x,Q^2)+g_2(x,Q^2)
\Bigr ] dx.   \nonumber
\end{eqnarray}

The basic assumptions leading to the dispersion relations between the 
forward Compton amplitudes and the generalized spin polarizabilities are the 
same as those leading to the generalized GDH sum rule.
However, since the generalized polarizabilities have an extra $1/\nu^2$ 
weighting compared to the GDH sum or 
 $I_3$, these integrals
converge
much faster, which minimizes the issue
of extrapolation to the unmeasured region at large $\nu$. For the kinematics
of this experiment, the contributions to the generalized polarizabilities 
from the unmeasured region are negligible. 

At low $Q^2$, the 
generalized polarizabilities have been evaluated with $\chi$PT 
calculations~\cite{KV02,BHM03}.
One issue in the $\chi$PT calculations is how to properly
include the nucleon resonance contributions, especially the $\Delta$ resonance,
which is usually dominating.
As was pointed out in Ref. \cite{KV02,BHM03}, while $\gamma_0$ is sensitive to 
resonances, $\delta_{LT}$ is insensitive to the $\Delta$ 
resonance. Measurements of the generalized spin
polarizabilities will be an important step in understanding the dynamics of
QCD in the Chiral Perturbation region.

We will first focus on the low $Q^2$ region where the comparison with $\chi$PT
calculations is meaningful, and then show the complete data set from
$Q^2$ of 0.1 GeV$^2$ to 0.9 GeV$^2$.

\begin{figure}[ht!]
\begin{center}
\centerline{\includegraphics[scale=0.40, angle=0]{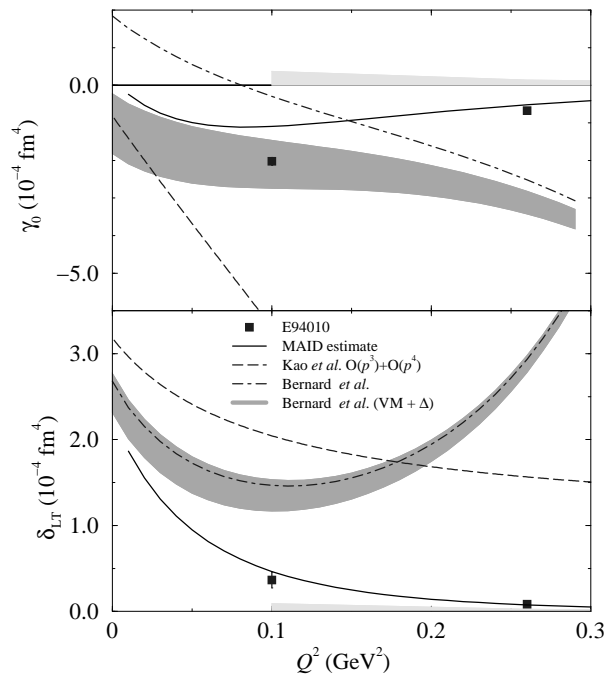}}
\end{center}
\caption{Forward spin polarizabilities $\gamma_0$ (top panel) and $\delta_{LT}$
(bottom panel). Solid squares are the results with statistical uncertainties.
The light bands are the systematic uncertainties. 
The dashed curves are the Heavy Baryon $\chi$PT 
calculation from Ref.~\cite{KV02}. 
The dot-dashed curves and the dark bands are the Relativistic 
Baryon
$\chi$PT calculation from Ref.~\cite{BHM03}, without and with 
the $\Delta$ and vector meson contributions, resepctively.
Solid curves are from the MAID model~\cite{maid}.
}
\label{fig:fig1}
\end{figure}

The results of $\gamma_0(Q^2)$ for the neutron are shown
in the top panel of Fig.~\ref{fig:fig1} as a function of $Q^2$ for the two 
lowest
 $Q^2$  values of 0.10 GeV$^2$ and 0.26 GeV$^2$. 
 The statistical uncertainties are generally smaller than the
size of the symbols. The systematic uncertainties are dominated by the uncertainties in the 
radiative corrections, the spectrometer acceptance and the beam and target 
polarization measurements. 
The data are compared with 
a next-to-leading order, $O(p^4)$, Heavy Baryon $\chi$PT 
(HB$\chi$PT)
calculation~\cite{KV02}, a next-to-leading-order
Relativistic Baryon $\chi$PT (RB$\chi$PT)
calculation~\cite{BHM03}, and the same calculation explicitly including 
both the $\Delta$ resonance and vector meson contributions.
Predictions from the MAID model~\cite{maid} are also shown.
At the lowest $Q^2$ point of 0.1 GeV$^2$
the RB$\chi$PT
calculation including the resonance contributions
is in good agreement with the experimental result.
For the HB$\chi$PT calculation without explicit resonance contributions, 
discrepancies are large even at $Q^2 = 0.1$ GeV$^2$. 
This might indicate the significance of the resonance contributions or a
problem with the heavy baryon approximation at this $Q^2$.
The higher $Q^2$ data point is in good agreement with the MAID
prediction,
but the lowest data point at $Q^2 = 0.1 $ GeV$^2$ is significantly lower, 
consistent with what was observed for the generalized GDH
integral results~\cite{Ama:02} and the underestimation from MAID for the 
neutron GDH sum rule at the real photon point~\cite{maid}.

Since the longitudinal-transverse spin polarizability $\delta_{LT}$ is 
insensitive to the dominating
$\Delta$ resonance contribution, it was believed that $\delta_{LT}$ should be
more suitable than $\gamma_0$ to serve as a testing ground for the chiral 
dynamics of QCD~\cite{BHM03,KV02}.
The bottom panel of Fig.~\ref{fig:fig1} shows $\delta_{LT}$ for the 
neutron compared to
$\chi$PT calculations and the MAID predictions. It is surprising to see
that 
the data are in significant disagreement with the $\chi$PT calculations 
even at the lowest $Q^2$, 0.1 GeV$^2$. 
This disagreement presents a significant challenge to the present theoretical
understanding. 
The MAID predictions are in good agreement with our results.

\begin{figure}[ht!]
\begin{center}
\centerline{\includegraphics[scale=0.4,angle=-90]{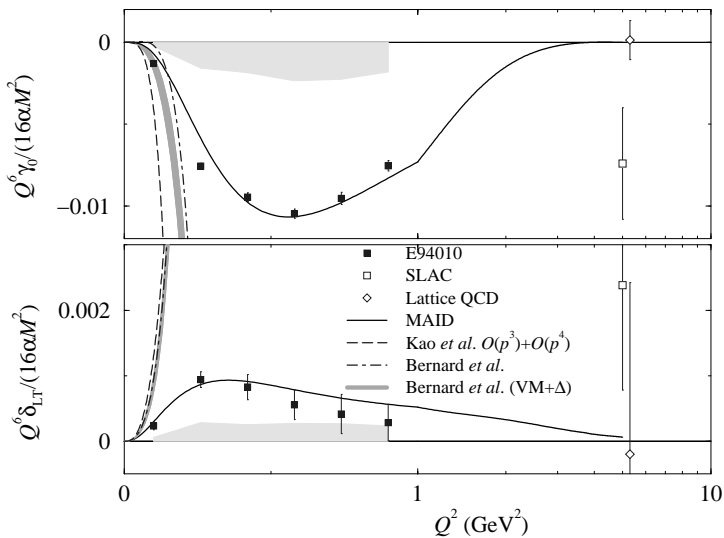}}
\end{center}
\caption{Forward spin polarizability 
$\gamma_0$ (top panel) and $\delta_{LT}$
(bottom panel) with $Q^6$ weighting. The solid squares are the
 results with statistical uncertainties.
The light bands are the systematic uncertainties. The open squares are the SLAC
data~\cite{E155} and the open diamonds are the Lattice QCD calculations~\cite{LQCD}. 
The curves are the same as in Fig. 1.}
\label{fig:fig2}
\end{figure}

Table 1 lists the experimental results for all $Q^2$ values.
Figure~\ref{fig:fig2} shows the results of both $\gamma_0$ and 
$\delta_{LT}$ multiplied by $Q^6$ along with 
the MAID and $\chi$PT calculations. Also shown are the world
data~\cite{E155} and a quenched Lattice QCD 
calculation~\cite{LQCD}, both at $Q^2= 5$~GeV$^2$.

\begin{table}%[H] add [H] placement to break table across pages
 \caption{\label{tab:table1}Results for $\gamma_0 (Q^2)$ and 
$\delta_{LT} (Q^2)$, along with statistical and systematic uncertainties}
 \begin{ruledtabular}
 \begin{tabular}{ccc}
$Q^2$ & $\gamma_0 \pm $ stat. $\pm$ syst. & $\delta_{LT} \pm $ stat. $\pm$ syst. \\
(GeV$^2$)&($10^{-4}$ fm$^4$)&($10^{-4}$ fm$^4$) \\
\hline
0.10 &$-2.02  \pm 0.11  \pm 0.36$  & $0.364 \pm 0.092 \pm 0.091$ \\
0.26 &$-0.67  \pm 0.015 \pm 0.14$  & $0.084 \pm 0.011 \pm 0.025$ \\
0.42 &$-0.200 \pm 0.005 \pm 0.039$ & $0.018 \pm 0.004 \pm 0.005$ \\
0.58 &$-0.084 \pm 0.002 \pm 0.019$ & $0.004 \pm 0.002 \pm 0.002$ \\
0.74 &$-0.037 \pm 0.001 \pm 0.009$ & $0.002 \pm 0.001 \pm 0.001$ \\
0.90 &$-0.016 \pm 0.001 \pm 0.004$ & $0.001 \pm 0.001 \pm 0.000$ \\
 \end{tabular}
 \end{ruledtabular}
 \end{table}

%\begin{table}%[H] add [H] placement to break table across pages
% \caption{\label{tab:table2}Results of $\delta_{LT} (Q^2)$ and uncertainties}
% \begin{ruledtabular}
% \begin{tabular}{cccc}
%$Q^2$ (GeV$^2$)& $\delta_{LT} (10^{-4} fm^4)$& stat.& syst.\\
%\hline
%0.10 &-2.023 &0.105 &0.364 \\
%0.26 &-0.673 &0.015 &0.140 \\
%0.42 &-0.200 &0.005 &0.039 \\
%0.58 &-0.084 &0.002 &0.019 \\
%0.74 &-0.037 &0.001 &0.009 \\
%0.90 &-0.016 &0.001 &0.004 \\
% \end{tabular}
% \end{ruledtabular}
% \end{table}

It is expected that
at large $Q^2$,
the $Q^6$-weighted spin polarizabilities become independent of $Q^2$
(scaling)~\cite{DPV}, and the deep-inelastic-scattering (DIS)
Wandzura-Wilczek 
relation~\cite{WW} leads to a relation 
between $\gamma_0$ and $\delta_{LT}$:

\begin{equation}
\delta_{LT}(Q^2) \rightarrow 
{\frac{1}{3}}\gamma_0(Q^2) \ \ \ {\rm as} \ \ Q^2 \rightarrow \infty. 
\end{equation}

\noindent 
  For inclusive DIS structure functions and their first moments the 
scaling behavior is observed to start around $Q^2$ of 1 GeV$^2$, 
where the higher twist effects become insignificant.
For the higher moments the scaling behavior is expected to start at a higher 
$Q^2$ than that for the first moments.
Our results show that 
scaling behavior is not observed at $Q^2 < 1$ GeV$^2$.  
 Again, both results are in good agreement with the MAID model.

In conclusion, we have made the first measurement of the forward spin 
polarizabilities $\gamma_0(Q^2)$ and  
$\delta_{LT}(Q^2)$ for the neutron in the $Q^2$ range from 0.1 GeV$^2$ to 0.9 GeV$^2$.
The low $Q^2$ results were compared to next-to-leading
order $\chi$PT calculations of two groups. The datum for
$\gamma_0$ at the lowest $Q^2$ is in good agreement  with the RB$\chi$PT
calculations including
explicit resonance contributions.
Although it
was expected that $\chi$PT calculations  should 
converge faster for $\delta_{LT}$ than for $\gamma_0$ 
as a result of smaller resonance contributions,
we find significant disagreement between data and both $\chi$PT 
calculations for $\delta_{LT}$.
The discrepancy presents a significant challenge to our theoretical understanding at 
its present level of approximations
 and might
indicate that higher
order calculations are needed for $Q^2 \ge 0.1$ GeV$^2$ 
and above.  
Our results, combining with future measurements at even lower $Q^2$
\cite{e97110}, will provide benchmark tests for our understanding of
QCD chiral dynamics. 
Except at the lowest $Q^2$ point for $\gamma_0$,
the rest of the new data agree well with the MAID model.
It shows that
the current level of phenomenological understanding of the resonance spin structure
in these observables is reasonable. In the $Q^2$ range of this experiment,
the expected high-$Q^2$ scaling behavior has not been observed yet. 

We would like to acknowledge the outstanding support of the Jefferson Lab
staff. We thank V. Bernard, D. Drechsel, T. Hemmert, B. Holstein, C. W. Kao, 
Ulf-G. Meissner, L. Tiator, M. Vanderhaeghen and their collaborators
for theoretical support and helpful discussions.
This work was supported by the U.S. Department of Energy (DOE), the U.S.
National  Science Foundation, the European INTAS Foundation, the Italian 
INFN and the French
CEA, CNRS, and  Conseil R\'egional d'Auvergne. The Southeastern Universities
Research Association operates the Thomas Jefferson National Accelerator
Facility for the DOE under  contract DE-AC05-84ER40150.


\vskip .1truein

\begin{references}

% Please use the \bibitem command to create references.


\bibitem{RCS} V. Olmos de Leon {\em et al.}, Eur. Phys. J. {\bf A 10}, 207 (2001); J. Tonnison {\em et al.}, Phys. Rev. Lett. {\bf 80}, 4382 (1998).

\bibitem{mainz} J. Ahrens {\em et al.}, Phys. Rev. Lett. {\bf 87}, 022003 (2001).

\bibitem{GPT} P. A. M. Guichon, G. Q. Liu and A. W. Thomas, Nucl. Phys.
{\bf A 591}, 606 (1995).

\bibitem{DPV} D. Drechsel, B. Pasquini and M. Vanderhaeghen, Phys. Rep. 
{\bf 378}, 99 (2003).

\bibitem{GPE} J. Roche {\em et al.}, Phys. Rev. Lett. {\bf 85}, 708 (2000).

\bibitem{ciofi} C. Ciofi degli Atti and S. Scopetta, Phys. Lett. B {\bf 404},
223 (1997).

\bibitem{Ama:02} M. Amarian {\em et al.}, Phys. Rev. Lett. {\bf 89}, 242301 (2002).

\bibitem{Ama:03} M. Amarian {\em et al.}, Phys. Rev. Lett. {\bf 92}, 022301 (2004) 

\bibitem{GDH} S. B.~Gerasimov, Sov.\ J.\ Nucl.\ Phys.\ {\bf 2}, 598 (1965);
S. D. Drell and A. C.~Hearn, Phys.\ Rev.\ Lett.\  {\bf 16}, 908 (1966).

\bibitem{GGDH1} D. Drechsel, S. S. Kamalov and L. Tiator, Phys. Rev.
{\bf D 63}, 114010 (2001).

\bibitem{GGDH2} X. Ji and J. Osborne, J. of Phys. {\bf G 27}, 127 (2001).

\bibitem{KV02} C.~W.~Kao, T. Spitzenberg and M. Vanderhaeghen, Phys. Rev. {\bf D  67}, 016001 (2003).

\bibitem{BHM03} V. Bernard, T. R. Hemmert and Ulf-G.  Meissner, Phys. Rev. 
{\bf D 67}, 076008 (2003).


\bibitem{maid} D. Drechsel, S. S. Kamalov and L. Tiator, Phys. Rev. {\bf D 63}, 114010 (2001).

\bibitem{E155} P. L. Anthony, {\em et al.}, Phys. Lett. {\bf B 493}, 19 (2000);
 {\it ibid.} {\bf 553}, 18 (2003).

\bibitem{LQCD} M. Gockeler, {\em et al.}, Phys. Rev. {\bf D 63}, 074506, (2001).
\bibitem{WW} S. Wandzura and F. Wilczek, Phys. Lett.  {\bf B 72}, 195 (1977).

\bibitem{e97110} Jefferson Lab E97-110, J.~-P.~Chen, A.~Deur and F.~Garibaldi, spokespersons;
\url{http://hallaweb.jlab.org/experiment/E97-110/}.



\end{references}
\end{document}